# A More Accurate Model for Finding Tutorial Segments Explaining APIs

He Jiang[1,2,3] jianghe@dlut.edu.cn
Jingxuan Zhang[1] jingxuanzhang@mail.dlut.edu.cn
Xiaochen Li[1] li1989@mail.dlut.edu.cn
Zhilei Ren[1] zren@dlut.edu.cn
David Lo[4] davidlo@smu.edu.sg

[1]School of Software, Dalian University of Technology, Dalian, China
[2]Key Laboratory for Ubiquitous Network and Service Software of Liaoning Province, Dalian, China
[3]State Key Laboratory of Software Engineering, Wuhan University, Wuhan, China
[4]School of Information System, Singapore Management University, Singapore

*Abstract*—**Developers prefer to utilize third-party libraries when they implement some functionalities and Application Programming Interfaces (APIs) are frequently used by them. Facing an unfamiliar API, developers tend to consult tutorials as learning resources. Unfortunately, the segments explaining a specific API scatter across tutorials. Hence, it remains a challenging issue to find the relevant segments. In this study, we propose a more accurate model to find the exact tutorial fragments explaining APIs. This new model consists of a text classifier with domain specific features. More specifically, we discover two important indicators to complement traditional text based features, namely co-occurrence APIs and knowledge based API extensions. In addition, we incorporate Word2Vec, a semantic similarity metric to enhance the new model. Extensive experiments over two publicly available tutorial datasets show that our new model could find up to 90% fragments explaining APIs and improve the state-of-the-art model by up to 30% in terms of F-measure.**

## I. INTRODUCTION

Instead of writing code snippets themselves, developers usually tend to reuse existing Application Programming Interfaces (APIs) to accomplish programming tasks [1, 2]. API anchored programming is a typical software reuse technique which can speed up developing process [3]. Unfortunately, it is a challenge for developers to use an unfamiliar API correctly, thus API learning resources are welcomed for developers. To be able to use an unfamiliar API without mistakes, developers usually tend to refer to the API specifications as learning resources [4, 5]. API specifications explain the preconditions, parameters, exceptions and return values of the specific APIs [6]. However, they will not put the specific APIs into some particular context. In contrast, tutorials are important learning resources which combine the functional explanations and code examples together. By referring to tutorials, developers can quickly know how to use the APIs correctly in specific situations. As a result, tutorials are key resources to the developers.

However, there are some challenges for developers to find the API explanations in the tutorials. First, tutorials are usually structured as a series of programming topics with APIs mixed in different sections. As a result, it is hard to find related sections for a given API in the tutorials. Developers often have to go through the lengthy tutorials to locate API explanations. Thus, it is essential to split the whole tutorials into consistent fragments in terms of content. Second, APIs are often mentioned in an irrelevant fragment without being the topic of it. One fragment may contain more than one API, but the fragment cannot explain all the appeared APIs at the same time. Developers have to decide whether the fragment is really talking about the given API, since the API may be listed as an example or appear to compare against other APIs. In such situations, the fragment does not explain this API.

Previous works have tried to discover tutorial fragments explaining APIs using a text classification model [1]. This model extracts a series of linguistic and structural features from the API-fragment pairs. These features are divided into five different groups according to their relevance, including real valued features, tutorial level, section level, sentence level features, and dependency based features. Real valued features detect the frequency of occurrence of the whole API and part of it. Tutorial level, section level, and sentence level features are all boolean or binary features which detect a predefined hypothesis. Dependency based features measure the dependency between the code-like terms. These features do not measure the similarity between APIs and fragments, since the information is unmatched. APIs usually consist of several terms, while the fragments contain hundreds of terms. After the features of each API-fragment pair are extracted, and the class label, namely whether the fragment explains the API or not is annotated, the training set is fed into the MaxEnt classifier to train the model. Given a new API-fragment pair, the trained model can predict the class label for it. The model can discover the fragments explaining APIs with an average F-measure of 75.6%.

However, the existing model does not consider the domain specific knowledge of APIs effectively. A given API is usually introduced along with other APIs which is called co-occurrence APIs, and the patterns of co-occurrence APIs are important factors to find explaining fragments for the given API. Besides, the given API can be treated as both text and code. From the perspective of text, we can fully mine its implicit meaning and relevance with the fragments. From the perspective of code, we can borrow the knowledge and API usage experience from both the crowds and experts, and these knowledge and experience can be used to find fragments explaining the given API. In summary, we have the following observations:

- Several APIs can appear in one fragment at the same time. When considering one of the APIs, the more co-occurrence APIs exist, the less possibility the fragment explains the API. As a result, the co-occurrence APIs are important indicators to discover fragments explaining an API.

- As a general programming interface, an API can be used by a large number of developers. The API usage experience can be discussed and shared in some technical forums, such as Stack Overflow. The crowd knowledge contained in Stack Overflow can be leveraged to discover explaining fragments for APIs.

- Accompanied with the distribution of APIs, the providers will supply the specifications for the APIs. The specifications are written by some experienced developers, and describe the functional properties of the APIs. The expert knowledge contained in the specifications can also be exploited.

In this study, we propose a new model of FInding Tutorial Segments Explaining APIs (FITSEA). FITSEA fully leverages the above mentioned domain specific knowledge, and follows the same process as text classification. In FITSEA, we construct three feature groups from each API-fragment pair. The first group is raw API feature group, and it digs relevant information between the raw APIs and fragments. The second group is co-occurrence API feature group. This group considers the structural and semantic information of co-occurrence APIs in the fragments. The last group is API extension feature group, and it takes advantages of the crowd knowledge and expert knowledge from Stack Overflow and API specifications respectively to extend APIs, and finds connections between API extensions and fragments. Within the three groups of features, we introduce the Word2Vec [28] similarity which can measure the semantic similarity and eliminate the mismatch between APIs and fragments. After all the features and label of each API-fragment are constructed, a widely used classifier, namely decision tree is utilized to train the classification model. The trained model can be used to predict the labels of new API-fragment pairs.

Experiment results over the dataset created by Gayane et al. [1] show that, FITSEA can achieve an average F-measure of 82.82% which outperforms the state-of-the-art model by 7.2%. To generalize the effectiveness of FITSEA, we annotate new dataset of Android API tutorials. Results on the newly annotated dataset also show the advantages of FITSEA model. For example, the F-measure is between 60% and 74% over the new dataset which is superior to the state-of-the-art model.

In summary, this study makes the following contributions:

- We propose a new model, namely FITSEA to discover the tutorial fragments explaining APIs. It fully leverages domain specific knowledge, and incorporates the Word2Vec semantic similarity. This model shows better results than the state-of-the-art model.

- We conduct a series of experiments to show the effectiveness of FITSEA. The experiment results indicate that FITSEA can achieve the F-measure of 82.82% and 67.22% over the two tutorial datasets on average, and outperforms the state-of-the-art model by 7.2% and 12.2% respectively.

- We construct a new manually annotated tutorial dataset related to Android APIs. The new tutorial dataset consists of four tutorials including 430 API-fragment pairs in total. The new tutorial dataset is made publicly available to researchers and developers.

This paper is structured as follows. In section II, we present the usage scenario. In section III, we show the tutorial datasets used in the experiments. We describe the overall framework of FITSEA in section IV. The experimental setup and results are introduced in section V and VI. Then, section VII and VIII show the threats to validity and related work respectively. Last, we conclude this study in section IX.

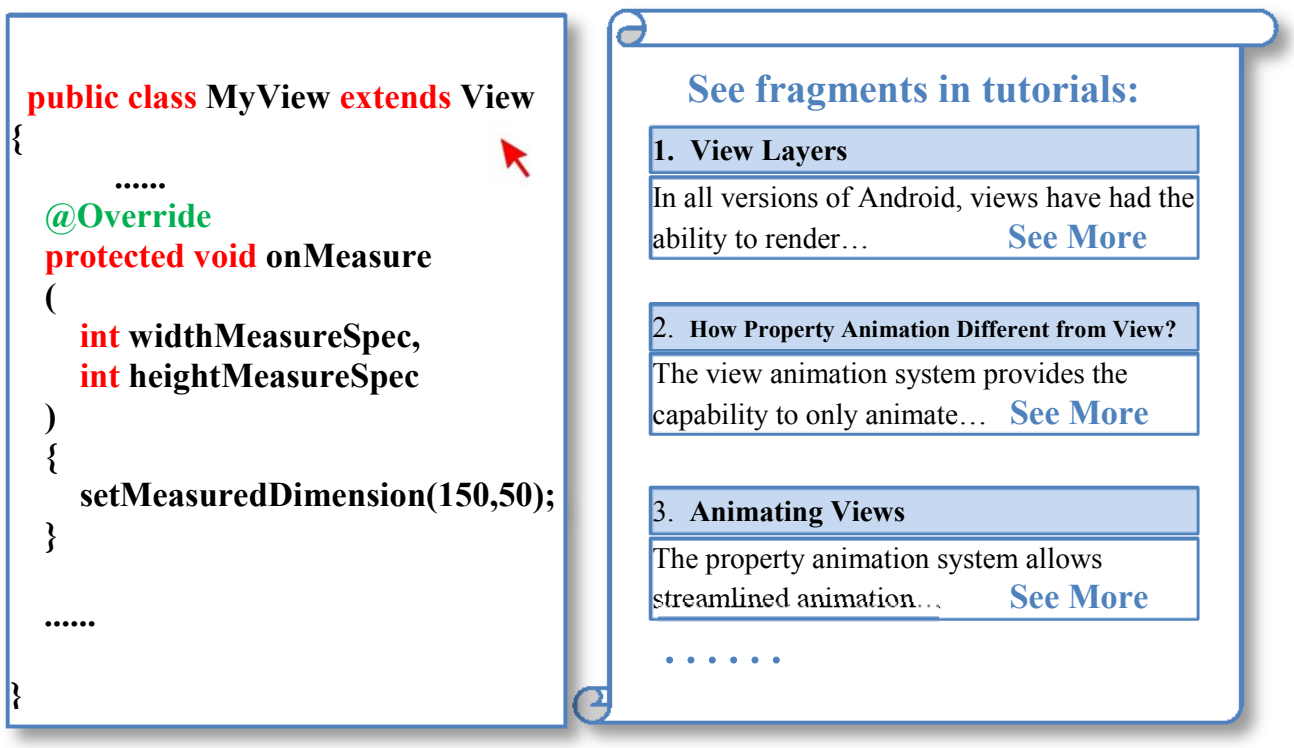

Fig. 1 An example of usage scenario

## II. USAGE SCENARIO

In this section, we demonstrate the usage scenario of discovering tutorial fragments explaining APIs.

When developers want to accomplish some programming tasks, they usually tend to reuse functionalities provided by third party libraries through APIs [21]. In such a situation, developers often know which specific APIs they should use, due to the meaningful name of the APIs. However, they often do not know how to use them in particular programming context. The aforementioned situation is of concern to us. Actually, there exists the situation which developers do not know which APIs they can use. However, many studies address the problem of API usage recommendation in the literature [7, 8, 9], and they can resolve this problem effectively. Hence, we do not take this situation into consideration.

For the above API usage scenario, we choose the granularity of the considered APIs at the class or interface level to discover the tutorial fragments as [1]. Tutorials usually introduce some programming topics by using a series of methods of some classes. Besides, developers usually want to know some behaviors of classes or interfaces rather than the functional feature of only one method [1]. As a result, Class or interface level is the best level of granularity.

By discovering the explaining tutorial fragments for an API, we can show the tutorial fragments to developers when they have no idea about the API. Fig.1 shows the usage scenario. If developers do not know how to use "android.view.View", we will recommend some fragments explaining it in the right panel. After reading the summary and the first sentence, developers can check more information by clicking "See More". In this way, developers will quickly learn how to use the API without much effort.

## III. Tutorial Datasets

It is necessary to find or construct tutorial datasets to study the process of discovering explaining fragments for APIs. There are two tutorial datasets used in this study. In the following subsections, we introduce the two tutorial datasets respectively, especially the process of constructing the Android tutorial dataset.

### *A. McGill Tutorial Dataset*

The first tutorial dataset denoted as McGill tutorial dataset is created by McGill University which is made publicly available. McGill tutorial dataset consists of five tutorials which explain JodaTime, Math, Collections and Smack APIs. They are diverse in the tutorial size and format, so they are well-suited for research study. Table 1 shows the statistical information of McGill tutorial dataset. We can see from the table that the number of API-fragment pairs ranges from 68 to 220, and the average length of fragments is less than 250 words. The relevant column shows the number of pairs in which the fragments really explain the APIs, and it is between 30 and 56.

### *B. Android Tutorial Dataset*

In order to contribute more tutorial datasets to research and development study, we construct another tutorial dataset explaining Android APIs, and name it as Android tutorial dataset. Android development is gradually booming these years, and developers are pouring into Android development. As a result, there is a high demand to discover explanatory tutorial fragments related to various Android APIs. This pushes us to build Android API tutorial dataset ourselves. As a complement for McGill tutorial dataset, Android tutorial dataset can be used to generalize the effectiveness of FITSEA model. We can find from Table 1 that there are four tutorials in Android tutorial dataset, namely Graphics, Resources, Text and Data for Android APIs. There is no big difference between McGill tutorial dataset and Android tutorial dataset in statistics, except that the lengths of tutorials in the Android dataset are much longer. The Android tutorial dataset is publicly available in the following website: http://oscar-lab.org/paper/API/.

### *C. Construction Steps*

There are mainly four steps to complete the construction of Android tutorial dataset, namely tutorial download, tutorial fragmentation, API identification and manual annotation. We will introduce each step in detail in the following paragraphs.

#### *1) Tutorial Download*

In the first step, we need to download the tutorials from the official Android development websites [33]. There are several tutorials in the websites, and we select four of them to crawl, namely Graphics, Resources, Text and Data. We choose these tutorials for the following reasons. First, they explain basic Android development topics which are relevant to many developers. Second, these four tutorials are easy to understand so that they can be manually annotated quickly and accurately. Third, they have different lengths and formats of fragments which can simulate different situations. We download the webpages of these tutorials for further processing.

Table 1. Statistical information of datasets

| Dataset | Tutorial | API | Fragment | Pairs | Length | Relevant |
|---|---|---|---|---|---|---|
| McGill Tutorial Dataset | JodaTime | 36 | 29 | 68 | 140 | 30 |
| | Math Library | 73 | 41 | 98 | 203 | 54 |
| | Col. Official | 59 | 57 | 220 | 172 | 56 |
| | Col. Jenkov | 28 | 69 | 150 | 141 | 42 |
| | Smack | 40 | 47 | 86 | 229 | 56 |
| Android Tutorial Dataset | Graphics | 70 | 38 | 138 | 411 | 43 |
| | Resources | 63 | 46 | 140 | 674 | 45 |
| | Text | 31 | 24 | 76 | 352 | 25 |
| | Data | 37 | 25 | 76 | 365 | 28 |

#### *2) Tutorial Fragmentation*

To help developers in finding useful information for an API quickly, we need to split the tutorials into short fragments. The contents in each fragment should be cohesive, so that they can concentrate on only one topic. Since the contents of the tutorials are all stored in HTML files, a basic idea is to split the tutorials based on HTML header tags. We find that the structures of these HTML files are the same, and the header tags are in four levels, namely <h1>, <h2>, <h3> and <h4>. Consequently, we decide to split the tutorials based on the lowest level, namely <h4>. By using regular expressions, we can split the tutorials.

#### *3) API Identification*

After splitting the tutorials into short fragments, we need to find the APIs in each fragment. Detecting the APIs is not a hard thing, since HTML files have already given the links to specific APIs. What we need do is to detect the "href" links in each fragment. By analyzing the link addresses, we can identify the exact APIs a fragment contains.

#### *4) Manual Annotation*

As we have already explained, not all the occurred APIs are explained by the fragment. After detecting the APIs in each fragment, we can combine them into API-fragment pairs. As a result, we need to annotate the class label of each pair by our subjective judgment.

We employ 6 master students to annotate Android tutorial dataset. They all major in software engineering, so it is not hard for them to annotate API related dataset. Before the annotation, each annotator is given a detailed annotation guide, and they are required to learn the guide until they know the whole annotation process. The annotation guide explains the aim of the annotation, the annotation procedures, the annotation standards and some tips to speed up the annotation. An example is also described in the annotation guide so that the annotators could get familiar with the annotation quickly.

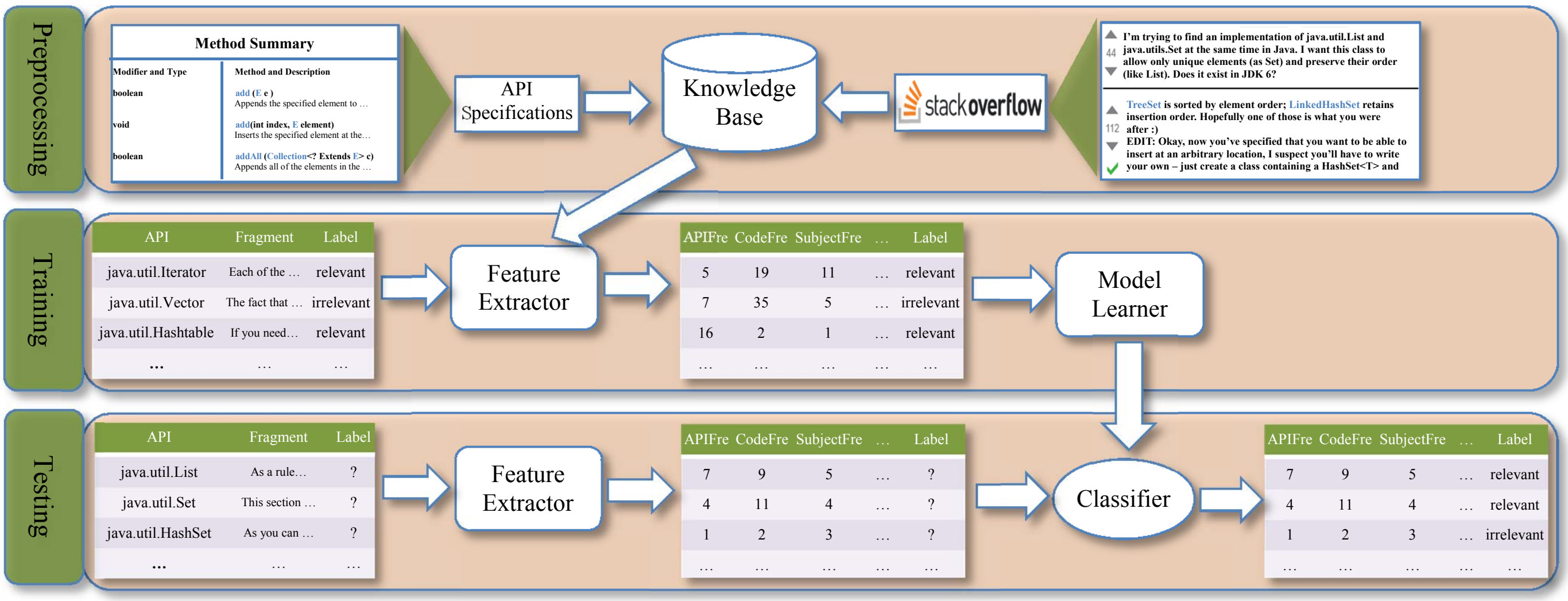

Fig. 2. Workflow of FITSEA

By reading an API-fragment pair, the annotators should decide whether the fragment explains the API. If yes, then the API-fragment pair is relevant, otherwise it is irrelevant. To make the annotation more reliable, each API-fragment pair is annotated by two different annotators. If there is a conflict between them when judging one API-fragment pair, they should discuss to reach a consensus. After the annotation process is finished, we collect the annotation results and construct Android tutorial dataset.

## IV. Framework

In this section, we illustrate FITSEA model, text similarity approaches, and the features that we use to characterize API-fragment pairs.

### A. *FITSEA Model*

FITSEA is a typical text classification model which is shown in Fig. 2. There are three stages in the whole framework. The first stage is the preprocessing stage which aims to construct the knowledge base from crowds and experts extracting from Stack Overflow and API specifications respectively. The second stage is the training stage which aims to transfer the API-fragment pairs into feature vectors with the assistance of a knowledge base and train a classifier. Each API-fragment pair has a class label, namely relevant or irrelevant. The third stage is the testing stage. After transforming the API-fragment pairs with unknown labels into feature vectors, the trained classifier can predict the class labels of the feature vectors in test set.

### B. *Text Similarity Approaches*

Since some text similarity methods are used as features or comparative methods in the experiments, we first describe the details of these methods. Texts can be similar in two ways, namely lexically and semantically [10]. Texts are similar lexically if they share a sequence of characters or terms (i.e., words), while texts are similar semantically if they show the same meaning. Lexical similarity can be further divided into character-based similarity and term-based similarity.

In this study, we utilize one of the semantic similarity methods, namely Word2Vec similarity. First, Word2Vec learns the vector representations of words. Then the vector representations of words in text can be merged together to form the final vector which can be treated as the semantic representation of the text. The similarity can be calculated based on the fixed-length vectors. In such a way, the mismatched length between two texts can be eliminated. To demonstrate the effectiveness of Word2Vec, we set up a research question to compare it against the other four lexical similarity methods, namely Bi-Gram, Levenshtein, Jaccard and Cosine similarity. Among them, Bi-Gram and Levenshtein are character-based similarity, while Jaccard and Cosine Similarity are term-based similarity. These text similarities can be calculated as follows.

#### 1) *Word2Vec Similarity*

Before calculating similarity, we first need to train the vector representation of each word. Word2Vec takes in a text corpus and outputs the vectors of all the words [34]. The text corpus is composed of all the words from the tutorial datasets, Stack Overflow, and API specifications. After the word vectors are obtained, we average the values in each dimension of each word vector in the text to form the text vector. The similarity between two text vectors is calculated as follows:

$$S(T1,T2)=\frac{\sum_{i=1}^{n} V(T1)_i V(T2)_i}{\sqrt{\sum_{i=1}^{n} V(T1)_i^2}\sqrt{\sum_{i=1}^{n} V(T2)_i^2}} \quad (1)$$

where T1 and T2 are two different texts, and $V(T1)_i$ and $V(T2)_i$ show the text vector of T1 and T2 at the ith dimension respectively.

#### 2) *Bi-Gram Similarity*

Bi-Gram is a type of N-Gram which splits the text into a sequence of characters with length 2 [11]. For example, the Bi-Gram of “student” is {st, tu, ud, de, en, nt}.The Bi-Gram similarity score is defined as the ratio of the number of shared Bi-Gram between two texts to the total number of Bi-Gram in both texts:

$$S(T1,T2)=\frac{2\times|Bi(T1)\cap Bi(T2)|}{|Bi(T1)|+|Bi(T2)|} \quad (2)$$

where Bi(T1) and Bi(T2) are the Bi-Gram of T1 and T2 respectively.

*3) Levenshtein Similarity*

Levenshtein similarity calculates the minimum edit operations needed to change one text into another. The edit operation includes insertion, deletion and substitution. The equation of it can be:

$$S(T1,T2)=1-\frac{MinOper(T1,T2)}{Max(T1,T2)} \quad (3)$$

where MinOper(T1,T2) stands for the number of minimum edit operations, and the Max(T1,T2) measures the maximum length of T1 and T2.

*4) Jaccard Similarity*

Jaccard similarity measures the size of intersection divided by the size of union in terms of words, and it can be calculated as follows:

$$S(T1,T2)=\frac{|S(T1)\cap S(T2)|}{|S(T1)\cup S(T2)|} \quad (4)$$

where S(T1) and S(T2) are the set of terms in T1 and T2.

*5) Cosine Similarity*

Texts can be transformed into word vectors, after a series of steps, such as tokenization, stemming and stop words removal. Cosine similarity measures the cosine of the angle between two word vectors, after transforming the texts into word vectors with the term weight of term frequency-inverted document frequency (tf-idf). The calculation formula can be:

$$S(T1,T2)=\cos(\theta)=\frac{\sum_{i=1}^{n} T1_i T2_i}{\sqrt{\sum_{i=1}^{n} T1_i^2}\sqrt{\sum_{i=1}^{n} T2_i^2}} \quad (5)$$

where $T1_i$ and $T2_i$ are word vectors of T1 and T2 at the ith dimension respectively.

### C. Feature Design

We design and extract 17 features from each API-fragment pair by taking co-occurrence APIs and API extensions into consideration. These features are divided into three groups, namely raw API features, co-occurrence API features, and API extension features respectively. Raw API features dig relevant information between raw APIs and fragments. Co-occurrence API features mine relationships between co-occurrence APIs and fragments. API extension features make use of the knowledge and usage experience from Stack Overflow and API specifications to find relevant facts. These groups of features all combine both linguistic and semantic properties between APIs and fragments. Some features have real values, while the others have boolean values. Table 2 shows the summary of these features. We will clarify all the features of each API-fragment pair in detail for the rest of this subsection.

*1) Raw API Features*

*WholeAPIFre:* This feature measures the frequency of the complete API name occurred in the fragment. The complete API name is the name from the root package, for example, the complete API name of "Iterator" is "java.util.Iterator". The rational is that the more times the complete API name appears, the more chances the fragment explains the API.

Table 2. Summary of features

| Feature | Description |
|---|---|
| **Group 1: Raw API features** | |
| WholeAPIFre | How strongly a whole API is associated to the fragment. |
| PartAPIFre | How strongly a part API is associated to the fragment. |
| ContainCodeFre | How many code snippets the fragment contains. |
| InstantiationFre | Frequency with which an API is initialized as an object. |
| SubjectFre | How many times the API acts as subject of each sentence. |
| InConditionSen | Whether the API appears in the condition sentences. |
| EmergeParaLoc | The minimum location the API appears in each paragraph. |
| Word2VecSimi | Word2Vec similarity between API and fragment. |
| **Group 2: Co-occurrence API features** | |
| CoAPIFre | How many co-occurrence APIs are contained in the fragment. |
| CoAPIFreInCode | How many co-occurrence APIs are contained in code snippets. |
| wholeCoAPIFre | How strongly co-occurrence APIs are associated to the fragment. |
| CoAPISenPro | The proportion of sentences which contain co-occurrence APIs. |
| CoWord2VecSimi | Word2Vec similarity between sentences containing co-occurrence APIs and not. |
| **Group 3: API extension features** | |
| MethodFre | How many methods in specification based API extension are contained in the fragment. |
| IsMethodInTitle | Whether the methods in specification based API extension are contained in the title. |
| ClueWordCount | How many clue words occurred in Stack Overflow based API extension. |
| ExWord2VecSimi | Word2Vec similarity between Stack Overflow based API extension and the fragment. |

*PartAPIFre:* An API name may be composed of several single words following CamelCasing convention. This feature measures the frequency of component words of API name in the fragment. For the same rational as the first feature, the more times the component words appear, the more chances the fragment explains the API.

*ContainCodeFre:* This feature computes code snippets frequency contained in the fragment. The code snippets can be found by the HTML tags, like "<codeblock>" or "<codebox>". The more code snippets exist in the fragment, the more chances the fragment concentrates on the API and its context.

*InstantiationFre:* This feature calculates how many times the API is initialized as an object. The more times it is initialized as an object, the more chances the fragment explains the constructors of the API.

*SubjectFre:* This feature calculates the frequency of which the API acts as the subjects of all the sentences. We use the Stanford Parser [12] to detect the subject of each sentence. If the API is the subject of a sentence, the sentence will pay much attention on it.

*InConditionSen:* This is a boolean feature which tries to detect whether the API exists in condition sentences. We detect the condition sentences with some phrases, such as "for example", "such as", "for instance" etc.

*EmergeParaLoc:* the minimum location where the API occurred in each paragraph of the fragment. If the API appears in the front of one paragraph, the paragraph will pay much attention on the API.

*Word2VecSimi:* The Word2Vec similarity between API name and fragment. The calculation method is introduced in part B of section IV.

*2) Co-occurrence API Features*

*CoAPIFre:* This feature detects the frequency of the co-occurrence APIs in the fragment in total. The more times the co-occurrence APIs appear, the less chances the fragment explains the API.

*CoAPIFreInCode:* The feature is calculated by counting the number of appearances of co-occurrence APIs in code snippets. The rational is the same as *CoAPIFre*.

*WholeCoAPIFre:* This feature combines both the complete co-occurrence APIs and part of them. If a complete co-occurrence API appears once, the value is increased by 1, while part of co-occurrence API (the component word) appears once, the value is increased by 0.5.

*CoAPISenPro:* This feature measures the proportion of sentences which contain co-occurrence APIs to all the sentences. The larger proportion of sentences containing co-occurrence, the less chance the fragment explains the API.

*CoWord2VecSimi:* The feature calculates the Word2Vec similarity between sentences which contain co-occurrence APIs and sentences do not.

*3) API Extension Features*

In the first place, we illustrate how to extend API from crowds and experts based on Stack Overflow and API specifications. Then, we show how to calculate features based on the API extensions.

Knowledge contained in Stack Overflow can be leveraged by many tasks [13, 14]. We follow the same method to discover crowd knowledge for APIs from Stack Overflow [15]. The APIs are treated as the queries, and the pairs of question and best answer are treated as documents to be retrieved. The ranking criteria are based on two aspects: the text similarity between the queries and the documents, and the quality of the documents. The text similarity is calculated using Lucene's [29] similarity, while the quality of the pairs of question and best answer can be measured by the user rating score of both question and best answer. The final ranking value is the average of text similarity and the quality of each pair of question and best answer. Through a series of experiments in [15], it has been proved to be an effective method for finding crowd knowledge from Stack Overflow. In FITSEA, we try to find crowd knowledge from Stack Overflow for APIs, and the first ranked pair of question and best answer is retained as Stack Overflow based API extension.

Besides, we parse the official API specifications, and extract the methods for each API. The specification based API extension consists of these methods. The following features are calculated using Stack Overflow based API extension or specification based API extension.

*MethodFre:* This feature calculates the frequency of the methods in specification based API extension appears in the fragment. The more times the methods exist, the higher chances the fragment explains the API.

*IsMethodInTitle:* This feature measures whether one of the methods in specification based API extension appears in the title of the fragment. If true, then the fragment will pay much attention on the API.

*ClueWordCount:* The clue words are the 10 highest frequent words [16] in the fragment. This feature measures how many clue words exist in the Stack Overflow based API extension.

*ExWord2VecSimi:* The feature calculates the Word2Vec similarity between the Stack Overflow based API extension and the fragment.

## V. Experimental Setup

In this section, we detail the experiment related issues. First, we show our Research Questions (RQs) to explore the performances of FITSEA. Second, we describe the two comparative models used in the experiments. Third, the evaluation method used in the experiments is introduced. Last, the evaluation metrics are explained.

### A. Research Questions

In this study, we investigate the following three RQs:

*RQ1: How will FITSEA perform when using different groups of features?*

As described above, features are divided into three groups. In this RQ, we want to explore the performances of FITSEA when applying different groups of features.

*RQ2: Does it achieve better results when using Word2Vec semantic similarity than other similarity methods?*

In each group of features, there exists one feature which measures semantic similarity based on Word2Vec. To explore whether it is superior to other similarity measurements, we try four other methods based on lexical information, namely Bi-Gram, Levenshtein, Jaccard and Cosine similarity.

*RQ3: Can FITSEA perform better than the other models over the two tutorial datasets?*

In this RQ, we want to explore whether FITSEA could discover more explaining fragments for APIs than the other models. We compare FITSEA against the state-of-the-art model which was introduced in [1] and an information retrieval model.

### B. Comparative Models

There are two comparative models in the literature [1]. The first model is proposed by [1] which we name it as GMR (constructed by concatenating the first character of each author's name). The second model is the traditional information retrieval model which we name it as IR. In the following part, we will detail the workflow of the two models.

*1) GMR Model*

The GMR model is a typical text classification model, and it is the first work which tries to discover explaining fragments for APIs. GMR model extracts 20 features from each API-fragment pair. The 20 features are linguistic and structural features, and they are divided into five groups. The training set is used to train a MaxEnt classifier, and the class labels of API-fragment pairs in test set can be predicted by the trained classifier.

*2) IR Model*

The IR model tries to resolve this task by using information retrieval method [27]. Since there is mismatch between information in APIs and fragments, the APIs should be extended using descriptions in the API specifications. The rational is that, the more conjoint words occurred in both API descriptions and fragment, the more likely the fragment concentrates on and explains the API. In such a situation, the API descriptions are treated as the queries, and the fragments are treated as the documents to be searched. The similarity is calculated using cosine similarity with tf-idf term weight.

We consider a fragment explains an API, if the cosine similarity between the fragment and the API descriptions is greater than a predefined threshold. The threshold is defined according to the following process: the top N fragments are received for each API, where N is the number of relevant fragments according to the annotation. The average value of each lowest similarity of the top N fragments is treated as the threshold. In such a way, IR model can also be evaluated in the same way as text classification model.

### C. Evaluation Method

In this subsection, we discuss the evaluation method when we perform text classification.

We use the same the evaluation method as in [1], namely Leave-One-Out Cross Validation (LOOCV) to test the results of each tutorial in all the experiments. More specifically, in each run, only one API-fragment pair is chosen as the test set. All the remaining pairs are treated as training set. Then, the pairs in training set are fed into the classifier to train the model. The trained model can be used to predict the class label of API-fragment pair in test set. After all the pairs are chosen as the test set, we calculate the final results. There are several advantages for using LOOCV. It could use as many pairs as the training set as possible, so the whole dataset can be effectively covered in each run. Besides, the results are more reliable and reproducible.

### D. Evaluation Metrics

In this study, widely used Precision, Recall and F-measure are employed to evaluate the performances of different models. Precision, Recall and F-measure are typical metrics to evaluate a classification model [16]. In the task of finding fragments explaining APIs, these metrics are also commonly used [1].

Table 3. Confusion matrix

| | | True Condition | |
|---|---|---|---|
| | | Positive | Negative |
| Predicted Condition | Positive | TP | FP |
| | Negative | FN | TN |

In the classification task, the True Positives (TP), True Negatives (TN), False Positives (FP) and False Negatives (FN) are computed to compare the results between true condition and predicted condition. Positive and negative refer to the prediction of the classifier, while true and false refer to whether the prediction corresponds to the true condition. The confusion matrix shown in Table 3 defines these terms.

Based on the confusion matrix, the Precision and Recall can be calculated as follows:

$$\text{Precision}=\frac{TP}{TP+FP} \tag{6}$$

$$\text{Recall}=\frac{TP}{TP+FN} \tag{7}$$

F-Measure tries to combine and balance the Precision and Recall which can be calculated as follows:

$$\text{F-Measure}=\frac{2\times\text{Precision}\times\text{Recall}}{\text{Precision}+\text{Recall}} \tag{8}$$

## VI. EXPERIMENTAL RESULTS

In this section, we detect the experimental results of each RQ. Through these RQs, we can evaluate different properties and performances of FITSEA.

### A. Investigation to RQ1

As mentioned before, we will investigate the results when using different groups of features. Different groups of features may have different contributions to the classifier. Through this RQ, we can also detect whether the co-occurrence API features and API extension features can improve the results effectively, as complements for the raw API features. Various combinations of the three groups are tested in each run. The results of each combination of groups of features over McGill dataset are listed in Table 4.

We can see from the table that different combinations of groups of features show different results over the five tutorials. For each separate group, the raw API feature group shows the best results on average. For example, the F-measure of JodaTime is 82.76% when using raw API feature group, while it's only 63.33% and 35.29% when only considering co-occurrence API features and API extension features respectively. For the combinations of two groups of features, the results are improved compared with only one group of features on average. For example, when combining raw API feature group with co-occurrence API feature group, the F-measure of Smack is 86.49%, while it's only 78.90% and 77.14% when considering one of them respectively. When the three groups of features are all considered, the results are the best on average. For instance, the F-measure of Col. Jenkov is 85.37% which is the best results. In summary, from the perspective of feature groups, there is no dominant or idle feature group. As more features are used, the results are improved on average.

Table 4. Classification results for different groups of features

| Group | JodaTime | | | Math Library | | | Col. Official | | | Col. Jenkov | | | Smack | | |
|---|---|---|---|---|---|---|---|---|---|---|---|---|---|---|---|
| | P (%) | R (%) | F-M (%) | P (%) | R (%) | F-M (%) | P (%) | R (%) | F-M (%) | P (%) | R (%) | F-M (%) | P (%) | R (%) | F-M (%) |
| 1 | 85.71 | 80.00 | 82.76 | 66.15 | 78.18 | 71.67 | 80.00 | 71.43 | 75.47 | 82.86 | 69.05 | 75.32 | 81.13 | 76.79 | 78.90 |
| 2 | 63.33 | 63.33 | 63.33 | 67.11 | 92.73 | 77.86 | 53.13 | 30.36 | 38.64 | 75.00 | 7.14 | 13.04 | 64.29 | 96.43 | 77.14 |
| 3 | 42.86 | 30.00 | 35.29 | 75.44 | 78.18 | 76.79 | 36.36 | 7.14 | 11.94 | 81.25 | 30.95 | 44.83 | 70.00 | 100.00 | 82.35 |
| 1+2 | 90.00 | 90.00 | 90.00 | 70.59 | 65.45 | 67.92 | 75.51 | 66.07 | 70.48 | 86.84 | 78.57 | 82.50 | 87.27 | 85.71 | 86.49 |
| 1+3 | 83.33 | 83.33 | 83.33 | 67.35 | 60.00 | 63.46 | 78.43 | 71.43 | 74.77 | 83.78 | 73.81 | 78.48 | 81.36 | 85.71 | 83.48 |
| 2+3 | 51.72 | 50.00 | 50.85 | 84.48 | 89.09 | 86.73 | 52.63 | 35.71 | 42.55 | 58.14 | 59.52 | 58.82 | 68.83 | 94.64 | 79.70 |
| 1+2+3 | 90.00 | 90.00 | 90.00 | 76.36 | 76.36 | 76.36 | 76.47 | 69.64 | 72.90 | 87.50 | 83.33 | 85.37 | 87.93 | 91.07 | 89.47 |

P: Precision, R: Recall, F-M: F-measure    1: raw API feature group, 2: co-occurrence API feature group, 3: API extension feature group

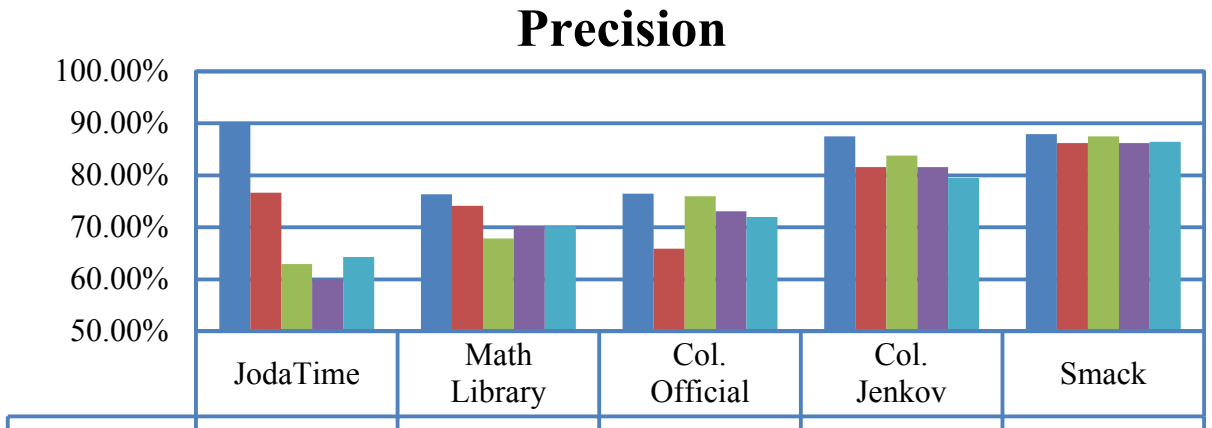

| | JodaTime | Math Library | Col. Official | Col. Jenkov | Smack |
|---|---|---|---|---|---|
| Word2Vec | 90.00% | 76.36% | 76.47% | 87.50% | 87.93% |
| VSM | 76.67% | 74.14% | 65.91% | 81.58% | 86.21% |
| Jaccard | 62.96% | 67.86% | 76.00% | 83.78% | 87.50% |
| Edit | 60.00% | 70.18% | 73.08% | 81.58% | 86.21% |
| Bi-Gram | 64.29% | 70.18% | 72.00% | 79.49% | 86.44% |

Fig. 3. Precision of different similarity methods

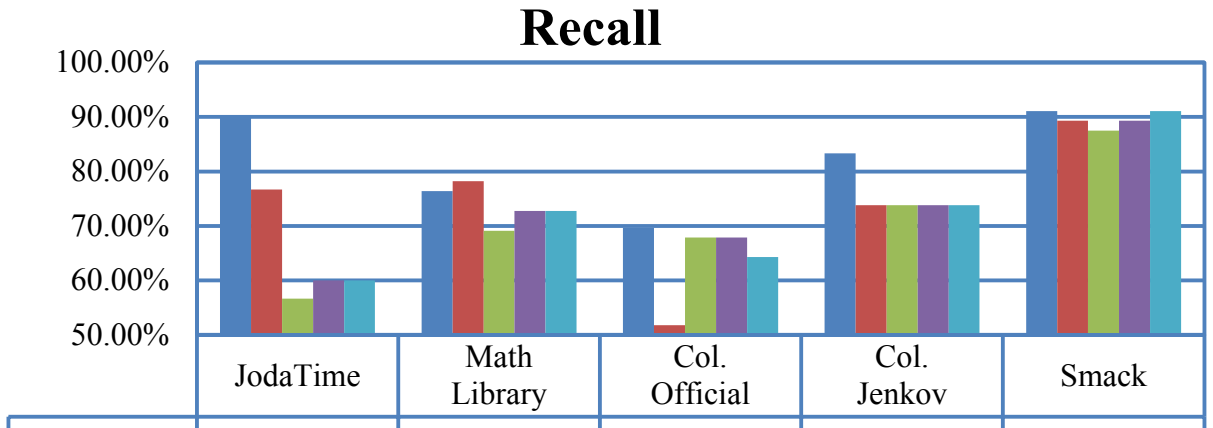

| | JodaTime | Math Library | Col. Official | Col. Jenkov | Smack |
|---|---|---|---|---|---|
| Word2Vec | 90.00% | 76.36% | 69.64% | 83.33% | 91.07% |
| VSM | 76.67% | 78.18% | 51.79% | 73.81% | 89.29% |
| Jaccard | 56.67% | 69.09% | 67.86% | 73.81% | 87.50% |
| Edit | 60.00% | 72.73% | 67.86% | 73.81% | 89.29% |
| Bi-Gram | 60.00% | 72.73% | 64.29% | 73.81% | 91.07% |

Fig. 4. Recall of different similarity methods

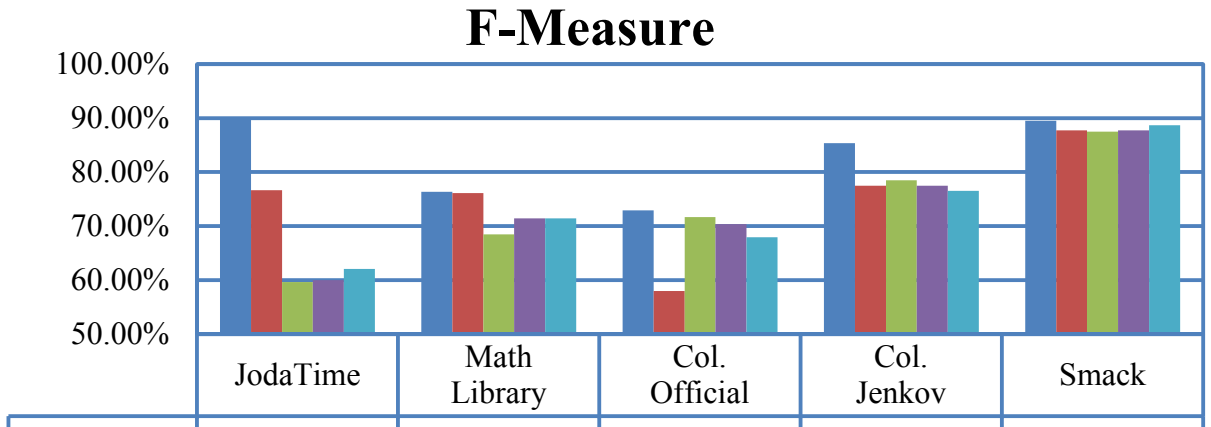

| | JodaTime | Math Library | Col. Official | Col. Jenkov | Smack |
|---|---|---|---|---|---|
| Word2Vec | 90.00% | 76.36% | 72.90% | 85.37% | 89.47% |
| VSM | 76.67% | 76.11% | 58.00% | 77.50% | 87.72% |
| Jaccard | 59.65% | 68.47% | 71.70% | 78.48% | 87.50% |
| Edit | 60.00% | 71.43% | 70.37% | 77.50% | 87.72% |
| Bi-Gram | 62.07% | 71.43% | 67.92% | 76.54% | 88.70% |

Fig. 5. F-measure of different similarity methods

When comparing different tutorials, we can see that even using the same feature group, different tutorials show different results. The F-measure is from 72.90% to 90% when using all the features among these tutorials. Col. Official is the most difficult tutorial to classify, the reason may be that the number of relevant and irrelevant API-fragment pairs is the most imbalance one. While JodaTime is the easiest tutorial to classify, since the number of relevant and irrelevant API-fragment pairs is the most balance one. We can see that, the more the data is balanced, the better the results are.

**Answer to RQ1:** With the increasing of the feature groups, the results get better on average. Co-occurrence API feature group and API extension feature group are good indicators for classification, and they are good complements for raw API feature group.

### B. *Investigation to RQ2*

When designing some features, we introduce Word2Vec to calculate the similarity. We design this RQ to test whether it is a more effective method than the others. In this RQ, we try the other four similarity methods to compare, namely Bi-Gram, Levenshtein, Jaccard and Cosine similarity. We only replace the features which calculate Word2Vec similarity with one of these methods, and the other features stay the same. Fig. 3, 4 and 5 show the Precision, Recall and F-measure of each tutorial over McGill tutorial dataset.

We can see that, the best results are achieved when using Word2Vec similarity. For example, the F-measure is 90% when using Word2Vec, while it is only 59.65% when using Jaccard in JodaTime tutorial. In other cases, Word2Vec also achieves better results than the other four similarity methods except for the Recall value in Math Library using VSM. By observing the results in different tutorials, we can find that, even one similarity method shows different effects. The F-measure cross all the tutorials is from 72.9% to 90% when using Word2Vec which shows the best results. The reason may be that, Word2Vec is a similarity method which can calculate the semantic similarity rather than lexical similarity. Even using different words, Word2Vec can capture the same meaning behind them. Besides, Word2Vec can learn fix-length vector representations of words, thus it can overcome the shortcomings of information mismatch.

**Answer to RQ2:** As a semantic similarity method, Word2Vec can measure the similarity much better than other methods. Word2Vec may be effective when the length of two texts is unmatched.

### C. *Investigation to RQ3*

In this RQ, we try to compare FITSEA with the other two

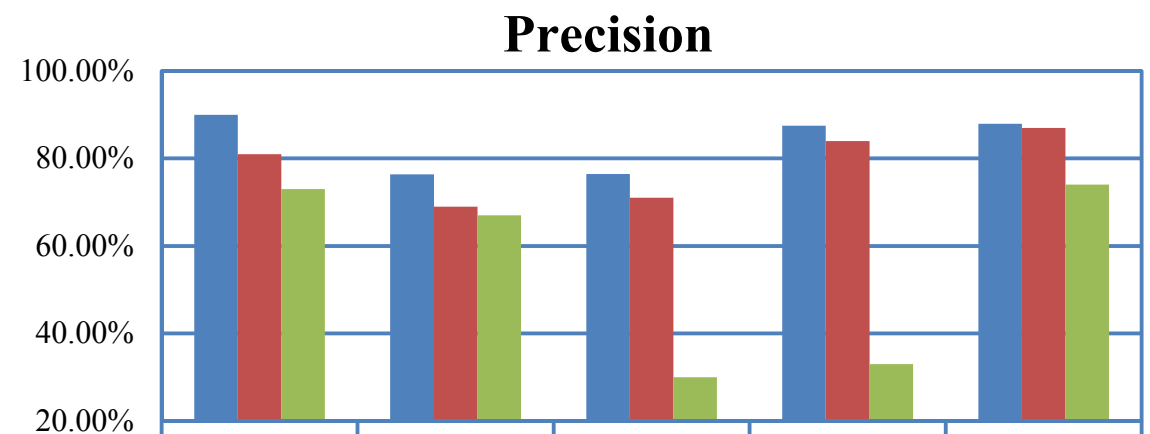

| | JodaTime | Math Library | Col. Official | Col. Jenkov | Smack |
|---|---|---|---|---|---|
| FITSEA | 90.00% | 76.36% | 76.47% | 87.50% | 87.93% |
| GMR | 81.00% | 69.00% | 71.00% | 84.00% | 87.00% |
| IR | 73.00% | 67.00% | 30.00% | 33.00% | 74.00% |

Fig. 6. Precision on McGill tutorial dataset

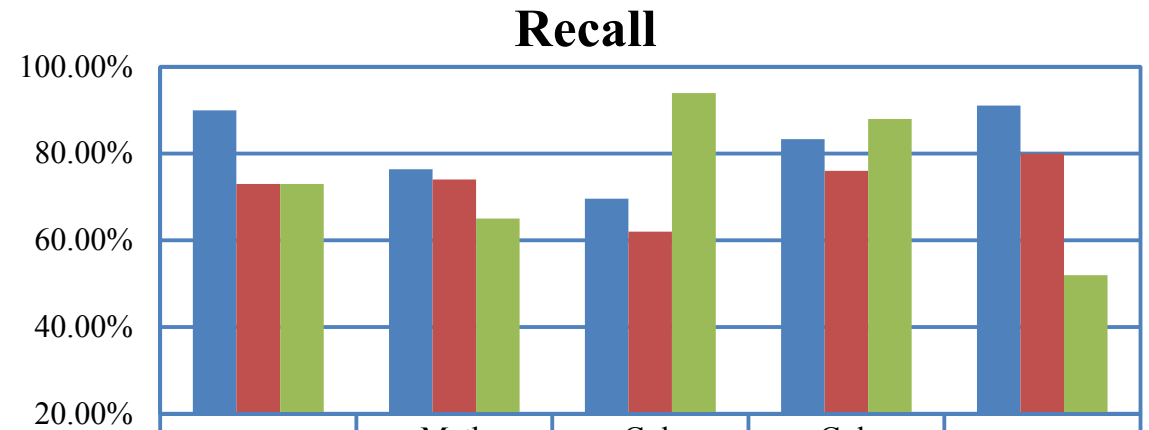

| | JodaTime | Math Library | Col. Official | Col. Jenkov | Smack |
|---|---|---|---|---|---|
| FITSEA | 90.00% | 76.36% | 69.64% | 83.33% | 91.07% |
| GMR | 73.00% | 74.00% | 62.00% | 76.00% | 80.00% |
| IR | 73.00% | 65.00% | 94.00% | 88.00% | 52.00% |

Fig. 7. Recall on McGill tutorial dataset

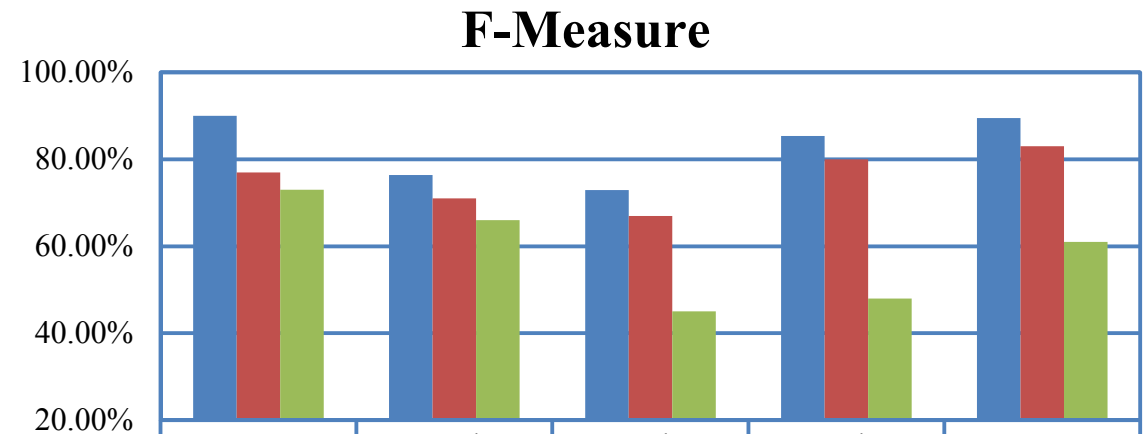

| | JodaTime | Math Library | Col. Official | Col. Jenkov | Smack |
|---|---|---|---|---|---|
| FITSEA | 90.00% | 76.36% | 72.90% | 85.37% | 89.47% |
| GMR | 77.00% | 71.00% | 67.00% | 80.00% | 83.00% |
| IR | 73.00% | 66.00% | 45.00% | 48.00% | 61.00% |

Fig. 8. F-measure on McGill tutorial dataset

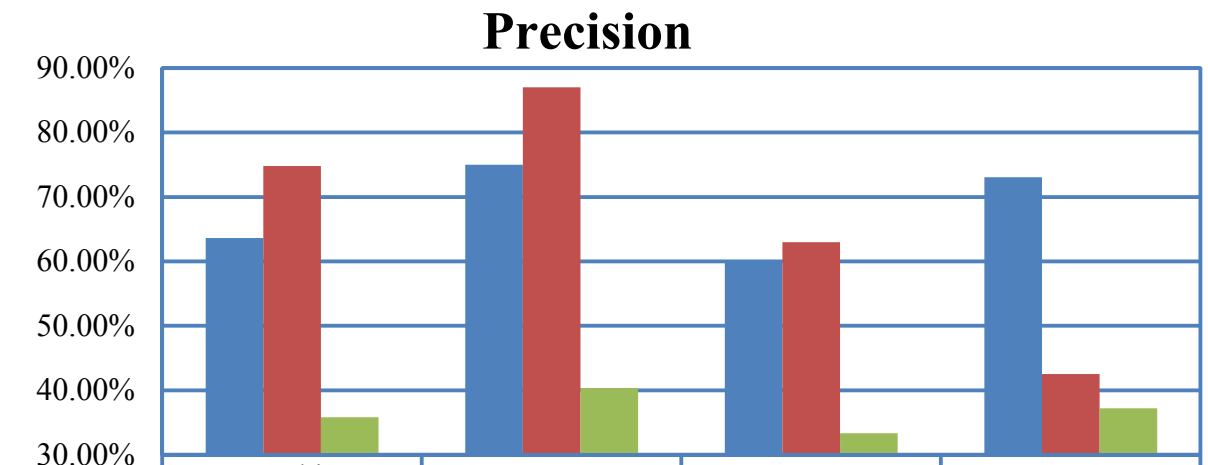

| | Graphics | Resources | Text | Data |
|---|---|---|---|---|
| FITSEA | 63.64% | 75.00% | 60.00% | 73.08% |
| GMR | 74.80% | 87.00% | 63.00% | 42.50% |
| IR | 35.80% | 40.32% | 33.33% | 37.21% |

Fig. 9. Precision on Android tutorial dataset

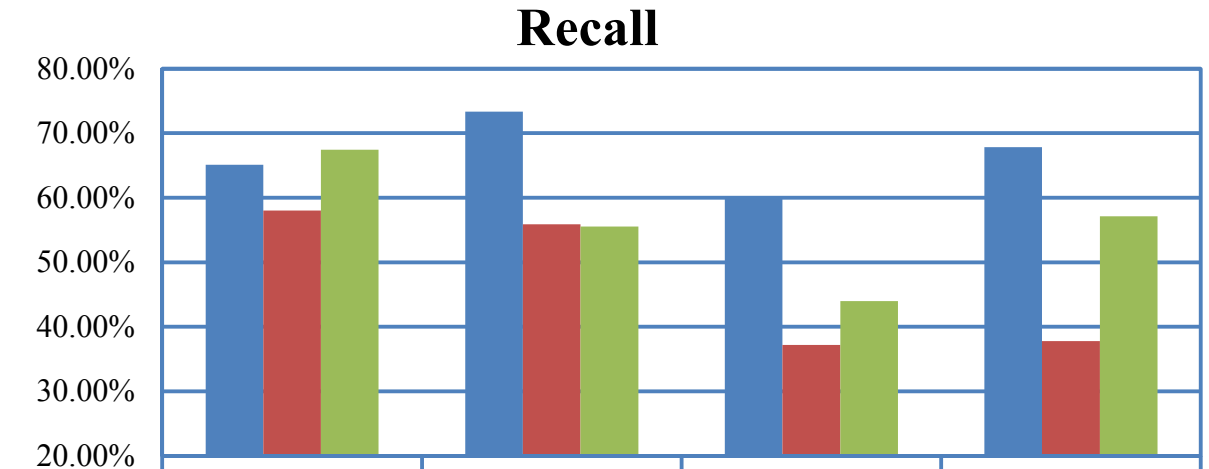

| | Graphics | Resources | Text | Data |
|---|---|---|---|---|
| FITSEA | 65.12% | 73.33% | 60.00% | 67.86% |
| GMR | 58.00% | 55.90% | 37.20% | 37.80% |
| IR | 67.44% | 55.56% | 44.00% | 57.14% |

Fig. 10. Recall on Android tutorial dataset

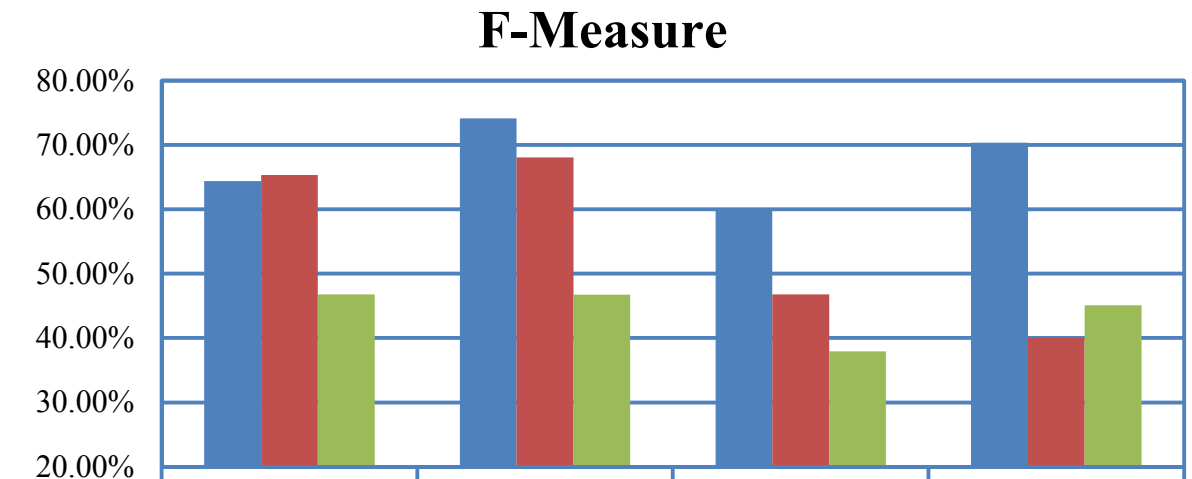

| | Graphics | Resources | Text | Data |
|---|---|---|---|---|
| FITSEA | 64.37% | 74.16% | 60.00% | 70.37% |
| GMR | 65.34% | 68.07% | 46.78% | 40.01% |
| IR | 46.77% | 46.73% | 37.93% | 45.07% |

Fig. 11. F-measure on Android tutorial dataset

models, namely GMR model and IR model. To show the effectiveness of FITSEA, we compare these models over both McGill tutorial dataset and Android tutorial dataset.

*1) Results on McGill Tutorial Dataset*

According to [1], GMR model is the state-of-the-art model which shows better results than IR model over McGill tutorial dataset. Fig 6, 7 and 8 show the Precision, Recall and F-measure respectively over McGill tutorial dataset.

We can see from the figures that, FITSEA shows better results on average than the two comparative models. In terms of Precision, FITSEA can achieve up to 90%, while GMR and IR can only reach up to 87% and 74% respectively. As for Recall, IR achieves better results than GMR while poorer than FITSEA. For example, we can see that IR can come to 94% on Col. Official. This is because the threshold is defined much smaller in IR model. As a result, a lot of fragments are retrieved. From the F-measure viewpoint, we can see FITSEA is better than the other two models on the whole.

*2) Results on Android Tutorial Dataset*

To make more contributions to the researchers and developers, we annotate the Android tutorial dataset. To demonstrate the generalization ability of FITSEA, we also compare FITSEA with GMR model and IR model over the Android tutorial dataset. Fig 9, 10 and 11 show the values of the three evaluation metrics over Android tutorial dataset.

We can see from the figures that FITSEA outperforms the other two comparative models. FITSEA doesn't achieve better results than GMR in some cases in Precision, but it shows better results than IR. In terms of Recall, FITSEA performs favorably to the comparative models. IR shows better results than GMR while poorer results than FITSEA on average. In particular, when it comes to F-measure, FITSEA reaches up to 74% and beyond GMR by up to 30% which shows absolute advantages.

After we have demonstrated the effectiveness of FITSEA, we would stress the underlying reasons why FITSEA works. It fully leverages the domain specific knowledge of APIs. Not only it mines the knowledge from Stack Overflow which can be called crowd knowledge, but also it borrows the expert knowledge from API specifications. Besides, through data inspection, we find that the more co-occurrence APIs exist, the less likely the fragment explains the API. Based on the observations, we design two groups of features which can improve the results. Last but not least, more advanced technologies have been used to help us, for example, Word2Vec semantic similarity.

**Answer to RQ3:** FITSEA shows better results than the state-of-the-art model over the two public datasets. FITSEA can find tutorial fragments explaining APIs more accurately.

## VII. THREATS TO VALIDITY

In this section, we discuss the threats to validity which include threats to internal validity and external validity.

### *A. Threats to Internal Validity*

In order to contribute more dataset used for discovering tutorial fragments explaining APIs and generalize the effectiveness of FITSEA, we employ annotators to create Android tutorial dataset. The annotation largely depends on the subjective judgment. Different annotators may have different viewpoints to the same thing due to their different backgrounds of computer and programming. To make it more reliable, the annotators are given a rigorous training. They are given a detailed guidance with examples to show the annotation process, criterions and so on. Besides, each API-fragment pair is annotated by two different annotators. If they have disagreements, they are required to discuss to reach a consensus. We believe that the above measures can eliminate the bias to some extent.

### *B. Threats to External Validity*

In this study, we employ two public open datasets to test the performances of FITSEA. It is still uncertain how FITSEA will perform on other tutorial datasets. Since the features can be defined and calculated accurately, FITSEA can show stable performances over different tutorial datasets. In the future, we plan to introduce more tutorial datasets to generalize FITSEA.

## VIII. RELATED WORK

In this section, we introduce two main related works, namely API usage patterns mining and text summarization.

### *A. API Usage Patterns Mining*

In addition to the recent studies in API documentation and discovery, APIs are hard to be used [18, 19, 20]. An API method is usually used along with the other methods to implement specific functionality. The patterns of co-usage relationships between methods of APIs are important for developers. Saied et al. [21] proposed a new method to mine multi-level API usage patterns. The usage patterns are created in a hierarchical way, and can be used by a variety of API client programs. Wang et al. [22] proposed an approach named UP-Miner which can find the unpopular usage patterns, and effectively reduce the number of redundant usage patterns.

API usage patterns can be used for different purposes, for example, API usage visualization [23, 24] and API usage example recommendation or code completion [25, 26]. When the patterns are discovered, the API usage patterns can be visualized to enhance understanding. If the database of API usage patterns can be constructed, they can be recommended to complete code snippets automatically.

### *B. Text Summarization*

This study relies heavily on natural language processing and text classification techniques. Finding tutorial fragments explaining APIs is similar to text summarization, especially extractive text summarization [16]. Extractive text summarization techniques have been used to many kinds of texts in software engineering, including meetings [30], telephone conversations [31] and bug reports [16, 32]. Software tasks around bug reports are popular topics these years [35, 36, 37]. Taking bug reports as examples, extractive text summarization can be resolved by supervised methods and unsupervised methods. Supervised methods train a classifier with various types of features to learn whether a sentence belongs to summary or not. Rastkar et al. [16] design 24 features to characterize sentences in bug reports. Time and location information are taken into account in these features. Unsupervised methods try to find summary without using a trained model. Mani et al. [32] compare four unsupervised methods for bug summarization, and introduce a noise reducer to filter out different types of sentences.

## IX. CONCLUSION

Finding tutorial fragments explaining APIs is significant to the developers which will speed up the development process. In this study, we propose a more accurate model, namely FITSEA to help developers finding tutorial fragments when facing an unfamiliar API. FITSEA fully leverages the domain specific knowledge to find two important indicators for classification, namely co-occurrence APIs and API extensions. Besides, it also introduces the usage of an effective semantic similarity method, namely Word2Vec. By investigating three RQs, we demonstrate the performances and effectiveness of FITSEA. The results show that FITSEA can outperform the state-of-the-art model by up to 13% and 30% on McGill tutorial dataset and Android tutorial dataset respectively in terms of F-measure.

## ACKNOWLEDGMENT

We greatly thank Petrosyan, Robillard, and Mori in McGill University for sharing their McGill tutorial dataset. We also thank the annotators who devote their time on annotating Android tutorial datasets. This work is supported in part by the National Program on Key Basic Research Project under Grant 2013CB035906, the New Century Excellent Talents in University under Grant NCET-13-0073, the National Natural Science Foundation of China under Grants 61370144 and 61403057, the Fundamental Research Funds for the Central Universities under Grant DUT14YQ203.